\documentclass[aps,floats,floatfix,amssymb,amsmath,prb,nofootinbib,twocolumn,superscriptaddress]{revtex4-2}

\usepackage{physics} 
\usepackage[T1]{fontenc}
\usepackage[latin1]{inputenc}
\usepackage{graphicx}
\usepackage{hyperref}
\usepackage{babel}
\usepackage{amsfonts}
\usepackage{color}
\usepackage{capt-of}
\usepackage{multirow}
\usepackage{xspace}
\usepackage[normalem]{ulem}

\newcommand{\wn}{\ensuremath{\rm cm^{-1}}\xspace}

\begin{document}
\title{Optical conductivity and vibrational spectra of the narrow-gap semiconductor FeGa$_3$} 

\author{C. Martin}
\affiliation{School of Theoretical and Applied Sciences, Ramapo College of New Jersey, Mahwah, New Jersey 07430, USA.}
\author{V. A. Martinez}
\affiliation{Department of Physics University of Florida, Gainesville, Florida 32611, USA}
\author{M. Opa\v ci\'c}
\affiliation{Institute of Physics Belgrade, University of Belgrade, Pregrevica 118, 11080 Belgrade, Serbia}
\author{S. Djurdji\'c-Mijin}
\affiliation{Institute of Physics Belgrade, University of Belgrade, Pregrevica 118, 11080 Belgrade, Serbia}
\author{P. Mitri\'c}
\affiliation{Institute of Physics Belgrade, University of Belgrade, Pregrevica 118, 11080 Belgrade, Serbia}
\author{A. Umi\'cevi\'c}
\affiliation{Vin\v ca Institute of Nuclear Sciences -- National Institute of the Republic of Serbia, University of Belgrade, P.O. Box 522, 11001 Belgrade, Serbia}
\author{A. Poudel}
 \affiliation{School of Theoretical and Applied Sciences, Ramapo College of New Jersey, Mahwah, New Jersey 07430, USA.}
\author{I. Sydoryk}
\affiliation{School of Theoretical and Applied Sciences, Ramapo College of New Jersey, Mahwah, New Jersey 07430, USA.}
\author{Weijun Ren}
\affiliation{Condensed Matter Physics and Materials Science Department, Brookhaven National Laboratory, Upton, New York 11973,  USA.}
\affiliation{Shenyang National Laboratory for Materials Science, Institute of Metal Research, Chinese Academy of Sciences,
Shenyang 110016, China}
\author{R. M. Martin}
\affiliation{Department of Physics and Astronomy, Montclair State University, Montclair, New Jersey 07043, USA}
\author{D. B. Tanner}
\affiliation{Department of Physics University of Florida, Gainesville, Florida 32611, USA}
\author{N. Lazarevi\'c}
\affiliation{Institute of Physics Belgrade, University of Belgrade, Pregrevica 118, 11080 Belgrade, Serbia}
\author{C. Petrovic}
\affiliation{Condensed Matter Physics and Materials Science Department, Brookhaven National Laboratory, Upton, New York 11973,  USA.}
\author{D. Tanaskovi\'c}
\affiliation{Institute of Physics Belgrade, University of Belgrade, Pregrevica 118, 11080 Belgrade, Serbia}

\begin{abstract}

Intermetallic narrow-gap semiconductors have been intensively explored due to their large thermoelectric power at low temperatures and a possible role of strong electronic correlations in their unusual thermodynamic and transport properties. Here we study the optical spectra and vibrational properties of $\mathrm{FeGa_3}$ single crystal. The optical conductivity indicates that $\mathrm{FeGa_3}$ has a direct band gap of $\sim 0.7$\,eV, consistent with density functional theory (DFT) calculations.
Most importantly, we find a substantial spectral weight also below 0.4~eV, which is the energy of the indirect (charge) gap found in resistivity measurements and ab initio calculations.
We find that the spectral weight below the gap decreases with increasing temperature, which indicates that it originates from the impurity states and not from the electronic correlations. 
Interestingly, we did not find any signatures of the impurity states in vibrational spectra.
The infrared and Raman vibrational lines are narrow and weakly temperature dependent. The vibrational frequencies are in excellent agreement with our DFT calculations, implying a modest role of electronic correlations. Narrow M\" ossbauer spectral lines also indicate high crystallinity of the sample.

\end{abstract}

\maketitle

\section{Introduction}

Correlated narrow-gap semiconductors represent a class of materials known for their large thermopower at low temperatures and other anomalous transport and thermodynamic properties \cite{Tomczak_review2018}. Three iron compounds among them, FeSi, FeSb$_{2}$ and FeGa$_{3}$, share some common features, but also show important differences. FeSi and FeSb$_{2}$ behave as insulators only at temperatures $T^* \lesssim 100 $ K which corresponds to the energy much smaller than the band gap $E_g \sim 50$~meV \cite{Schlesinger_FeSi1993,Park_1995,Petrovic_2003}. A build up of the in-gap spectral weight at tempe\-ratures $k_B T^* \ll E_g$, clearly seen in optical \cite{Damascelli_1997,Chernikov_1997,Perucchi_2006, Herzog_2010} and photoemission spectroscopy \cite{Arita_2008}, is a signature of strong electronic correlations \cite{Tomczak_FeSb2_2010,TomczakFeSi_2012}. A crossover from a non-magnetic insulator to a bad metal is accompanied by a large increase in the spin susceptibility which obtains Curie-Weiss form above room temperature \cite{Jaccarino_1967,Koyama_2007}. 
This crossover leaves fingerprints also in the Raman vibrational spectra which become strongly temperature dependent. The width of vibrational peaks increases several times in the bad-metal region as compared to the low-temperature insulating state \cite{Racu_2007,Lazarevic_2010,Lazarevic_2012}.
At temperatures near 10~K there is a large peak in the thermopower $|S|$ \cite{Sales_1994,Sun_2009}. The exact role of the electronic correlations, in-gap states, anisotropy and phonon-drag in colossal thermopower found in FeSb$_{2}$ remains a subject of various studies and controversy \cite{Homes2018,Battiato_PRL2015,Du2021}.

FeGa$_{3}$ has a significantly larger band gap, $E_{g}\sim$ 0.4 eV~\cite{Hadano2009,Wagner_Reetz_2014}, than FeSi and FeSb$_2$ due to the stronger hybridization between $3d$ orbitals of Fe and $4p$ orbitals of Ga. The electronic structure calculations imply modest contribution of electronic correlations. Density functional theory (DFT) \cite{Yin_2010} and LDA+U \cite{Wagner_Reetz_2014} calculations give almost the same band structure, while dynamical mean field theory (DFT+DMFT) \cite{Gamza_2014} gives only slightly reduced band gap. Nevertheless, the temperature dependence of dc resistivity is nontrivial: it strongly deviates from a simple activated transport at low temperatures, and features four distinct transport regimes which are associated with a presence of the in-gap states \cite{Hadano2009,Wagner_Reetz_2014,Gamza_2014}. For $T \lesssim 5$ K $\rho_{\mathrm{dc}}$ has a power law temperature dependence consistent with the variable-range hopping transport driven by the localized in-gap states. In the interval $20 \, \mathrm{K} \lesssim T \lesssim 45 \, \mathrm{K}$ the charge transport is activated, but corresponds to a small gap of $\sim 40$ meV between the in-gap states and the conduction band. Then, following a minimum in $\rho_{\mathrm{dc}}$, there is a metallic-like transport up to $\sim 80$ K which presumably corresponds to the regime where most of the in-gap electrons are already transferred to the conduction band. For $T> 300$ K the charge transport is activated, consistent with the wide gap $E_g \approx 0.4$ eV. 
The measurements show weak sample anisotropy and weak temperature dependence of magnetic susceptibility, whereas the DFT+DMFT calculations give small mass renormalization, as well as strong charge and spin fluctuations \cite{Gamza_2014}. 
A maximum in the Seebeck coefficient $|S|$ at $T\approx 15$ K is argued to be a consequence of the phonon-drag effect \cite{Wagner_Reetz_2014}. In this picture the in-gap states supply a free charge carriers and the acoustic phonons cause an additional scattering of the electrons opposite to the direction of a temperature gradient, leading to the large thermoelectric power. 
Interestingly, to our knowledge, there has been so far only one infrared spectroscopy study of $\mathrm{FeGa_3}$ in polycrystalline samples \cite{Knyazev_2017}, restricted to room temperature and energies larger than 90 meV.

In this paper, we present infrared and Raman spectroscopy study of $\mathrm{FeGa_3}$ single crystal in the temperature range between 4 and 300 K. The reflectance is measured in the energy interval between 30 and 24\,000 cm$^{-1}$. The infrared and Raman active vibrational frequencies are in excellent agreement with our DFT calculations, indicating good crystallinity and a small influence of electronic correlations. Good crystallinity is corroborated also by measured M\"ossbauer spectra. The most prominent feature of the optical spectra is the existence of the in-gap states below the charge gap of approximately 0.4~eV. We observe a reduction of the in-gap spectral weight as the temperature increases to 300~K and conclude that this spectral weight originates from the impurities. Details of experimental and numerical methods are presented in Sec.~II. The results are shown in Sec.~III and Sec.~IV contains our conclusions.

\section{Methods}

Single crystals of FeGa$_3$ were grown as described previously \cite{Hadano2009}. 
For infrared measurements a small crystal was polished until a smooth surface of about 3 mm$^{2}$ area was obtained, then mounted on a helium-flow optical cryostat. The temperature dependence of reflectance was measured between 30 and 24\,000~cm$^{-1}$, using a combination of two Fourier-transform infrared spectrometers: a Bruker 113v for far infrared (30 - 600~cm$^{-1}$) and a Bruker Vertex 70, with extended spectral range, from mid-infrared to visible (100 - 24\,000 cm$^{-1}$). Reflectance in visible and ultraviolet (12\,000 - 50\,000 cm$^{-1}$) was measured at room temperature only, using a Perkin-Elmer 650 UV/VIS grating spectrometer. As no temperature dependence was observed above about 12\,000 cm$^{-1}$, all temperatures were merged with room temperature data in visible and ultraviolet part of the spectrum. To capture correctly the width and line-shape of lattice vibrations, the far-IR data was taken with a resolution of 0.5 cm$^{-1}$, while 2~cm$^{-1}$ or larger values were used at higher frequencies.
Both gold and aluminum mirrors were used for reference, and in order to correct for surface roughness, the sample was also gold coated, using a commercial Ted Pella Cressington 108 sputtering machine. Because of the polishing involved, the precise orientation of the electric field (polarization) with respect to the crystallographic axes of the samples is not clearly defined, hence we cannot discuss potential anisotropic optical properties.

Raman scattering measurements were performed using a TriVista557 Raman system, equipped with a nitrogen-cooled CCD detector, in backscattering micro-Raman configuration. Grating configuration was 1800/1800/2400 grooves/mm, in order to achieve the best possible resolution. The 514.5~nm line of an Ar+/Kr+ gas laser was used as an excitation source and a microscope objective with factor 50 magnification was used for focusing the beam. All measurements were carried out with laser power less than 1.5~mW at the sample, in order to minimize local heating. Room temperature measurements were done in air, whereas for low temperature measurements the sample was placed in a KONTI CryoVac continuous flow cryostat, with a 0.5~mm thick window. Spectra were corrected for the Bose factor. 

The $^{57}$Fe-M\"ossbauer spectrum of the FeGa$_3$ powdered sample was measured at room temperature in high ($\sim \pm 9$~mm/s) and low ($\sim \pm 2$~mm/s) Doppler velocity range. The spectra were collected in standard transmission geometry in constant acceleration mode using a $^{57}$Co(Rh) source. The Doppler velocity scale was calibrated by using the M\"ossbauer spectrum of metallic $\alpha$-Fe. The spectra were fitted by the Recoil program \cite{Lagarec_1998}. The center shift value ($CS$) is quoted relative to the $\alpha$-Fe ($CS = 0$).

First-principles DFT calculations of electronic structure and phonon frequencies were performed using the open-source QUANTUM-ESPRESSO package \cite{Giannozzi_2009,Giannozzi_2017}. We employed the ultrasoft Vanderbilt-type pseudopotentials with Perdew-Burke-Ernzerhof exchange and correlation functional. For the Fe atom we considered 3$s$, 3$p$, 3$d$ and 4$s$ as valence electrons (in total 16), while the Ga valence electrons were taken to be the electrons from 3$d$, 4$s$ and 4$p$ orbitals (in total 13). Thus, a minimum of 110 bands was needed to perform the calculations since we have four formula units per unit cell, but we nevertheless considered 128 bands, which is a very convenient number for parallelization purposes. The plane waves kinetic energy cutoff was set to 70 Ry, which proved to be sufficient for all our calculations. The ionic relaxation, self-consistent and normal modes calculations were performed using the Monkhorst-Pack scheme, with the k-mesh of 8$\times$8$\times$8, which corresponds to 75 $k$-points in the irreducible part of the Brillouin zone. On the other hand, the density of states (DOS) calculation requires a much larger number of $k$-points in order to be accurate, and hence we performed the non-self-consistent calculation with a $k$-mesh of 12$\times$12$\times$12 in order to calculate the DOS. We used density functional perturbation theory (DFPT) \cite{Baroni_RMP2001} in order to calculate the vibrational frequencies.

\section{Results}

We first present the band structure calculations. These results are known from the literature, but we nevertheless show them for completeness and in order to put into context the analysis of the experimental data that follow. Then we present optical, Raman and M\"ossbauer spectra.

\subsection{DFT band structure}
The semiconductor FeGa$_{3}$ belongs to the $P4_2/mnm$ space group and it has a tetragonal $P$-type lattice with lattice parameters $a$=6.2628(3)\r{A} and $c$=6.5546(5)\r{A} ~\cite{HAUSSERMANN200294}. 
In the DFT calculations we used the lattice parameters from the experiment and relaxed only the fractional coordinates of the atoms.
These coordinates, shown in Table~\ref{table:Atoms}, are only slightly adjusted from their measured values.

\begin{table}[ht]
\caption{Nonequivalent atomic positions from the DFT calculation.} 
\centering 
\begin{tabular}{c c c c c } 
\hline\hline 
Atom & $P$\={4}$n2$ & x & y & z\\ [0.5ex] 
\hline 
Fe & 4f & 0.34367 & 0.34367 & 0\\ 
Ga1 & 4c & 0 & 0.5 & 0\\
Ga2 & 8j & 0.15575 & 0.15575 & 0.26295 \\ [1ex] 
\hline 
\end{tabular}
\label{table:Atoms} 
\end{table}

\begin{figure}[h!]
\includegraphics[width=\columnwidth]{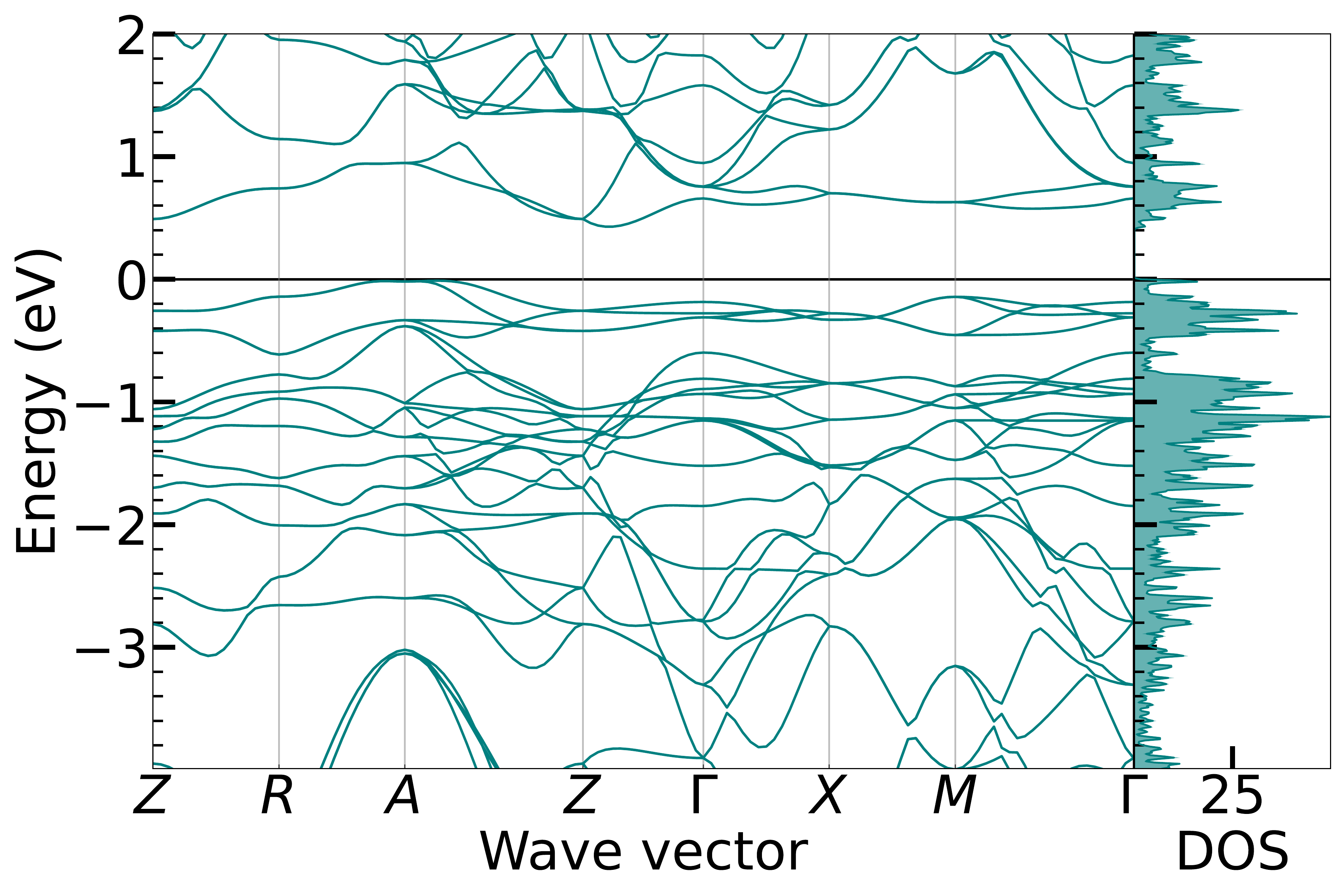}                
\caption{DFT band structure of FeGa$_3$ and density of states in units states/(eV\,f.u.).} 
\label{Fig:DFTDOS}
\end{figure}

Fig.~\ref{Fig:DFTDOS} shows the dispersion relations and the density of states, calculated along the $k$-path $Z-R-A-Z-\Gamma-X-M-\Gamma$ in the Brillouin zone. Our results are very similar to previous work~\cite{Wagner_Reetz_2014, IMAI2006722}, showing that FeGa$_{3}$ is an indirect-gap semiconductor with the calculated band gap of 0.44 $eV$. The bands around the Fermi level are formed from the hybridized Fe 3$d$ and Ga 4$p$ orbitals. 

\subsection{Optical conductivity and infrared vibrational modes}

\begin{figure}[t!]
\includegraphics[width=1.0\columnwidth]{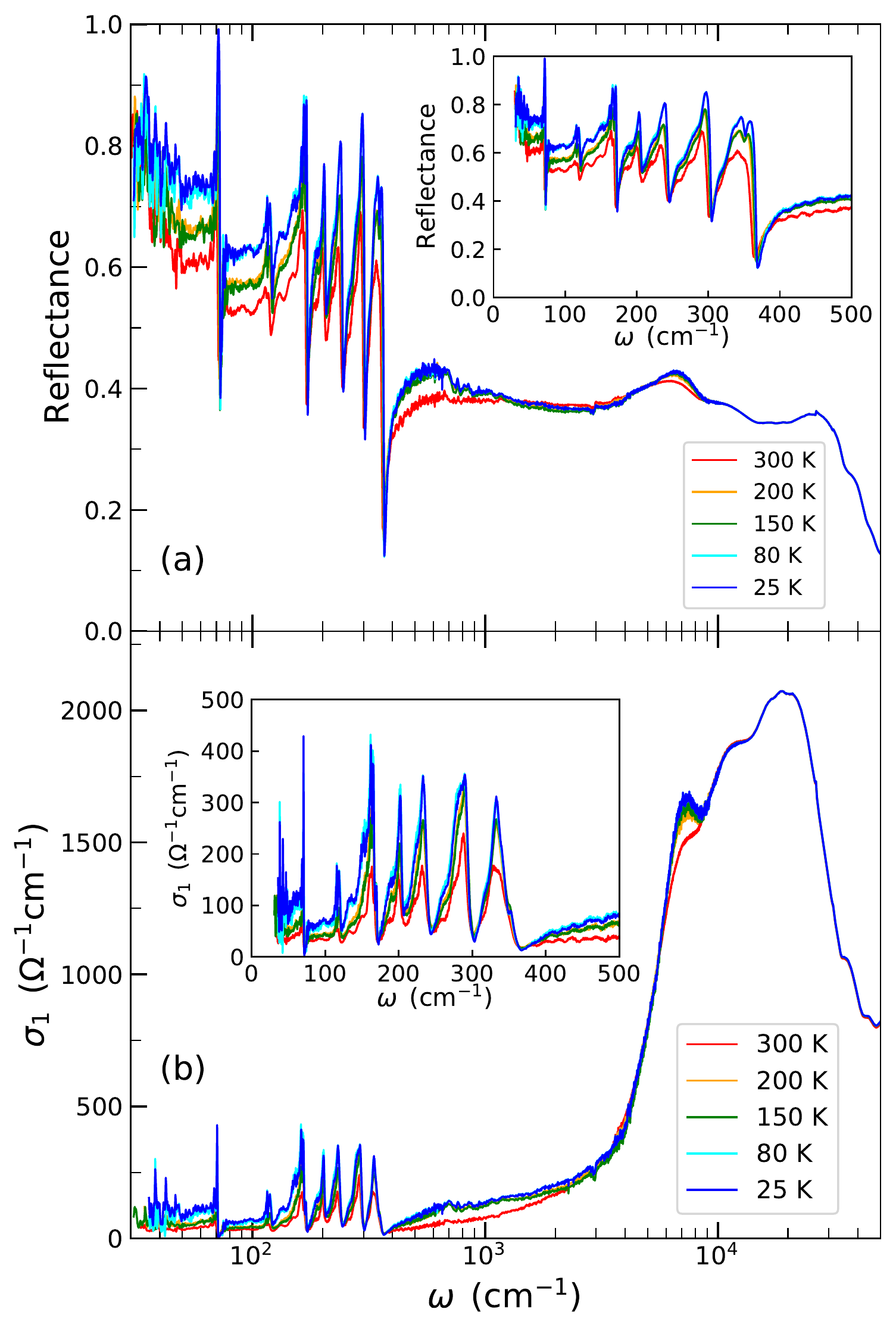}  
\caption{Reflectance (a) and optical conductivity (b) as a function of frequency in the whole measured frequency range at several temperatures. The insets show the low frequency data on a linear scale.} 
\label{Fig:IR1}
\end{figure}

The reflectance $R(\omega)$ measured at several temperatures between 25 and 300 K is shown in Fig.~\ref{Fig:IR1}(a). Note that the spectra are shown on a logarithmic frequency scale so that we can distinguish both the low- and high-frequency features. The low-frequency reflectance is close to 1 which indicates a possible presence of the in-gap states that we will discuss in detail below. The far-infrared frequency region is shown on a linear scale in the inset. The peaks in $R(\omega)$ correspond to the infrared-active vibrational modes.

A better insight into the excitation spectrum can be obtained from the real part of the optical conductivity $\sigma_1(\omega)$. It corresponds to the imaginary part of the dielectric function, $\sigma_{1}(\omega)=\omega\epsilon_{2}(\omega)/4\pi$, describing the absorption of electromagnetic radiation  \cite{FWooten,DBTanner}. 
Fig.~\ref{Fig:IR1}(b) shows 
$\sigma_{1}(\omega)$ obtained from the Kramers-Kronig transformation of $R(\omega)$. As this transformation involves integration of $R(\omega)$ from zero to infinity, we used extrapolation of our measurements. At high frequency ($\omega\rightarrow\infty$), the data were bridged with calculations of the dielectric function based on the X-ray photoabsorption, following the procedure described in Ref.~\cite{Tanner_2015}. For $\omega < 30 \, \mathrm{cm}^{-1}$ we set $R(\omega) = R(30 \, \mathrm{cm}^{-1})$, but we checked that $\sigma_1(\omega)$ is not sensitive to the precise form of $R(\omega)$ in the limit $\omega\rightarrow 0$. The same result is obtained using the Hagen-Rubens formula, $R(\omega)=1-A\sqrt{\omega}$, where $A$ is a constant adjusted to fit the first several points from the measurements~\cite{FWooten, DBTanner}.

\begin{figure}[t!]
\includegraphics[width=\columnwidth]{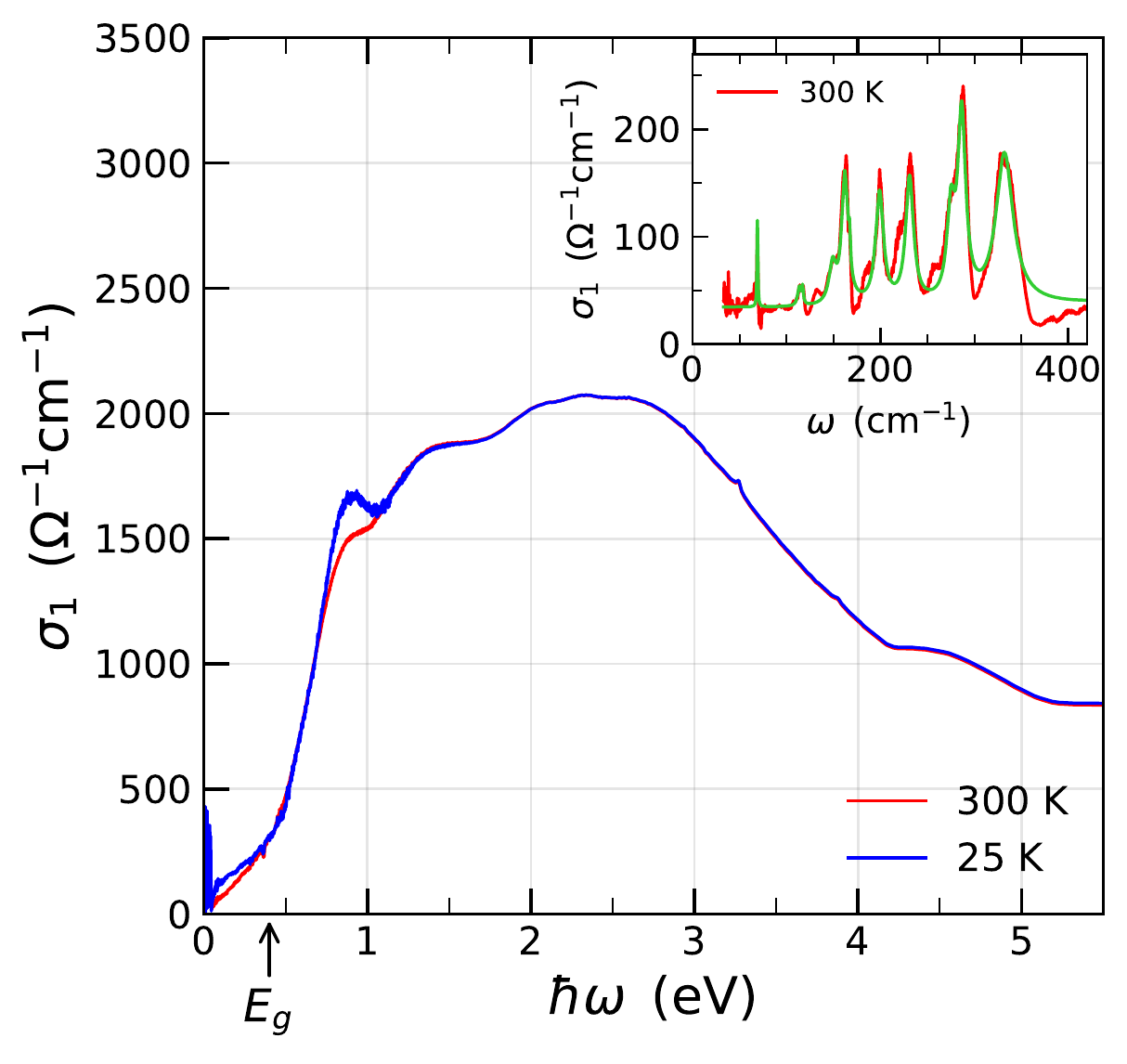}  
\caption{Optical conductivity as a function of frequency at 25 and 300 K. The inset shows the infrared vibrational modes which are fitted by 11 Lorentzians. The green line corresponds to the cumulative fit.}
\label{Fig:IR2}
\end{figure}

The optical conductivity at 25 K and 300 K is shown in Fig.~\ref{Fig:IR2} on a linear energy (frequency) scale in units of eV.
$\sigma_1(\omega)$ rapidly decreases for frequencies $\hbar \omega \lesssim  0.9 \, \mathrm{eV} \, (7\,000 \, \mathrm{cm}^{-1})$.
This is consistent with the DFT band structure shown in Fig.~\ref{Fig:DFTDOS}. It gives the smallest direct gap of 0.67~eV near the Z-point in the Brillouin zone, but in many regions of the Brillouin zone the gap is between 0.7 and 0.9~eV.
At $\hbar \omega = E_g \approx 0.4$~eV the spectral weight is significantly reduced, yet it remains substantial also at $\hbar \omega < E_g$. 
We note that we did not attempt to calculate the optical spectra since 
a reliable calculation requires to include the particle-hole interaction on a level of the Bethe-Salpeter equation, which is a very challenging task even for weakly interacting semiconductors \cite{Albrecht_BSE_1998,Rohlfing_BSE_2000}. A calculation of the optical spectra of FeGa$_{3}$ in the independent-particle approach poorly compares with our experiments \cite{Yin_2010}. On the other hand, our optical spectra for $\hbar \omega \gtrsim 0.5$ eV are in rather good agreement with the spectra on polycrystalline samples of Ref.~\cite{Knyazev_2017}.

Evidence that the spectral weight below $E_g$ has origins in
the impurity states
can be obtained from analysis of the temperature dependence of $\sigma_1(\omega)$. 
At finite temperatures, in a standard band gap semiconductor 
a small spectral weight would appear just below $E_g$ due to the phonon assisted excitations. The same amount of the spectral weight would recover just above the band edge, where the absorption becomes slightly lower due to the finite hole (electron) concentration in the valence (conduction) band at finite temperatures \cite{FWooten}. 
On the other hand, in FeSi and FeSb$_{2}$ a spectral gap is closed at higher temperatures due to the strong correlation effects. In this case, transfers of the spectral weight occur over the energy range much larger than the band gap.
However, in our case the spectral weight at $T=300 \, \mathrm{K}$ is reduced both below and above $E_g$, see Figs.~\ref{Fig:IR1}(b) and Fig.~\ref{Fig:IR2}. The reduction of the spectral weight below $E_g$ should correspond to the depopulation of the impurity band, which leads to the reduction in the light absorption for subgap energies \cite{Gamza_2014}. 
Hence, we conclude that the spectral weight below $E_g$ is due to impurity states. We note that a small surplus of Fe atoms in comparison to the stoichiometric ratio is found in wavelength dispersive x-ray spectroscopy \cite{Wagner_Reetz_2014}.
Our conclusion is in line with the statement that the transfer of the electrons from the impurity states to the conduction band is a likely cause of the anomalous $d\rho/dT > 0$ resistivity temperature dependence around 100 K \cite{Gamza_2014,Battiato_PRL2015}.

\begin{table}[!t]
\caption{Irreducible representation of infrared-active modes and their frequencies. The measured frequencies are obtained at 300\,K. Numerical values are obtained within DFPT calculation.
\vspace*{0.2cm}
}
\centering 
\begin{tabular}{l l l } 
\hline \hline 
Irred.~rep. & Exp. ($\mathrm{cm}^{-1}$) & Calc. ($\mathrm{cm}^{-1}$) \\ [0.5ex] 
\hline 
 $E_u^{1,2}$  & 69.00  & 76.08 \\
 $A_{2u}^1$  & 113.50 & 107.61  \\
 $E_u^{3,4}$  & 117.50 & 116.18 \\
 $E_u^{5,6}$  & 149.85 & 161.78 \\
 $A_{2u}^2$ & 162.2 & 162.28 \\
 $E_u^{7,8}$ & 166.99 & 168.69  \\
 $E_u^{9,10}$ & 199.50 & 201.71  \\
 $E_u^{11,12}$ & 231.50 & 229.45 \\
 $E_u^{13,14}$  & 275.00 & 281.2 \\
 $A_{2u}^3$ & 287.30 & 296.61   \\
 $E_u^{15,16}$ & 332.50 & 329.0  \\ [1ex]
\hline 
\end{tabular}
\label{IR_freq}
\end{table}

We now turn our attention to the far-infrared part of the spectrum from 50 to 350$~\mathrm{cm}^{-1}$, which contains infrared vibrational modes. From the inset of Fig.~\ref{Fig:IR1}(b) it appears that most of the phonon peaks are rather broad. However, that is not the case since several peaks, in fact, consist of two vibrational modes with very close frequencies.
The space group $P4_2/mnm$
has the corresponding point group $D_{4h}(4/mmm)$. Thus, all the normal modes are classified according to irreducible representations of $D_{4h}(4/mmm)$. The factor group analysis predicts 12 Raman and 11 infrared-active modes, along with ten silent and two acoustic modes: 
\begin{equation}
\label{eq:Phon}
\begin{aligned}
\Gamma_{\mathrm{Raman}} & =3A_{1g}+4E_{g}+2B_{1g}+3B_{2g} , \\
\Gamma_{\mathrm{IR}} & =3A_{2u}+8E_{u} ,\\
\Gamma_{\mathrm{silent}} & =2A_{2g}+2A_{2u}+4B_{1u}+2B_{2u},\\
\Gamma_{\mathrm{acoustic}} & =A_{2u}+E_{u}.
\end{aligned}
\end{equation}
%
The experimental data at 300 K are fitted with 11 Lorentz profile lines. Their cumulative contribution to the spectra is shown in green color in the inset of Fig.~\ref{Fig:IR2}. A complete list of the corresponding phonon frequencies is shown in Table~\ref{IR_freq}. These frequencies were obtained at 300\,K, but we see from the inset in Fig.~\ref{Fig:IR1}(b) that the temperature dependence of the frequencies is weak. The changes are of the order of 1\% in the temperature range between 25 and 300 K. The frequencies calculated within DFPT are in excellent agreement with measured frequencies. This implies that a crystallinity of the sample is very good, even though some surplus of Fe iron atoms is expected in comparison to the ideal composition \cite{Wagner_Reetz_2014}. In addition, excellent agreement between the calculated and measured frequencies indicates that the electronic correlations beyond the DFT are not strong, in line with the conclusions from DFT+U \cite{Wagner_Reetz_2014} and DFT+DMFT \cite{Gamza_2014} calculations.

\subsection{Raman spectra}

There are 12 Raman-active modes in FeGa$_3$, see Eq.~\ref{eq:Phon}. Wyckoff positions of the atoms, their contributions to the $\Gamma$-point phonons and the corresponding tensors for Raman active modes are given in Table~\ref{table1}.
Since observability of the Raman-active modes in backscattering configuration of the experiment strongly depends on the orientation of the sample, we first performed orientation dependent measurements. This was done by rotating the sample in the steps of 10$^\circ$. The orientation of the sample which provided the best observability of Raman modes of various symmetries was used in further measurements.

\begin{table}[b]
\caption{Contributions of each atom to the $\Gamma$-point phonons for the $P4_2/mnm$ space group and the corresponding tensors for Raman active modes.
\vspace*{0.2cm}
}
\label{table1}
\begin{ruledtabular}
\centering
\begin{tabular}{ccc}
\multicolumn{3}{c} {Space group $P4_2/mnm$} \\ \cline{1-3}  \\[-2mm]

Atoms &  \multicolumn{2}{c}{Irreducible representations} \\\cline{1-1} \cline{2-3} \\[-2mm]

\multirow{2}{*} {Fe ($4f$) }&  \multicolumn{2}{c}{$A_{1g} + A_{2g} + A_{2u}+ B_{1g}$} \\[1mm]
& \multicolumn{2}{c}{$+ B_{1u} + B_{2g} + E_{g} + 2E_u$}\\[1mm]

Ga ($4c$) &  \multicolumn{2}{c}{$A_{1u} + A_{2u} + B_{1u} + B_{2u} + 4E_u$}\\[1mm]

\multirow{2}{*}{Ga ($8j$)} & \multicolumn{2}{c}{$2A_{1g} + A_{1u} + A_{2g} + A_{2u}+ B_{1g}$}\\[1mm]

&\multicolumn{2}{c}{$+ 2B_{1u} + 2B_{2g} + B_{2u} + 3E_{g} + 3E_u$} \\[1mm]

\cline{1-3}  \\[-2mm]

$
A_{1g} = \begin{pmatrix}
a&0 &0 \\
 0&a&0 \\
 0&0 &b\\\end{pmatrix}
$
&
$B_{1g} = \begin{pmatrix}
c&0 &0 \\
 0&-c&0 \\
 0&0 &0\\
\end{pmatrix}
$&
$B_{2g} = \begin{pmatrix}
0& d&0 \\
d &0&0 \\
 0&0 &0\\
\end{pmatrix}
$\\

${}^1E_g = \begin{pmatrix}
c& 0 &0 \\
 0&0&e\\
 0& e&0 \\ \end{pmatrix}$
&&
$
{}^2E_g = \begin{pmatrix}
  0&0&-e\\
0&0  &0  \\
-e& 0&0  \\
\end{pmatrix}
$
\\ [5mm]

\end{tabular}
\end{ruledtabular}
\end{table}

\begin{figure}[h]
 \centering
  \includegraphics[width=7.2cm]{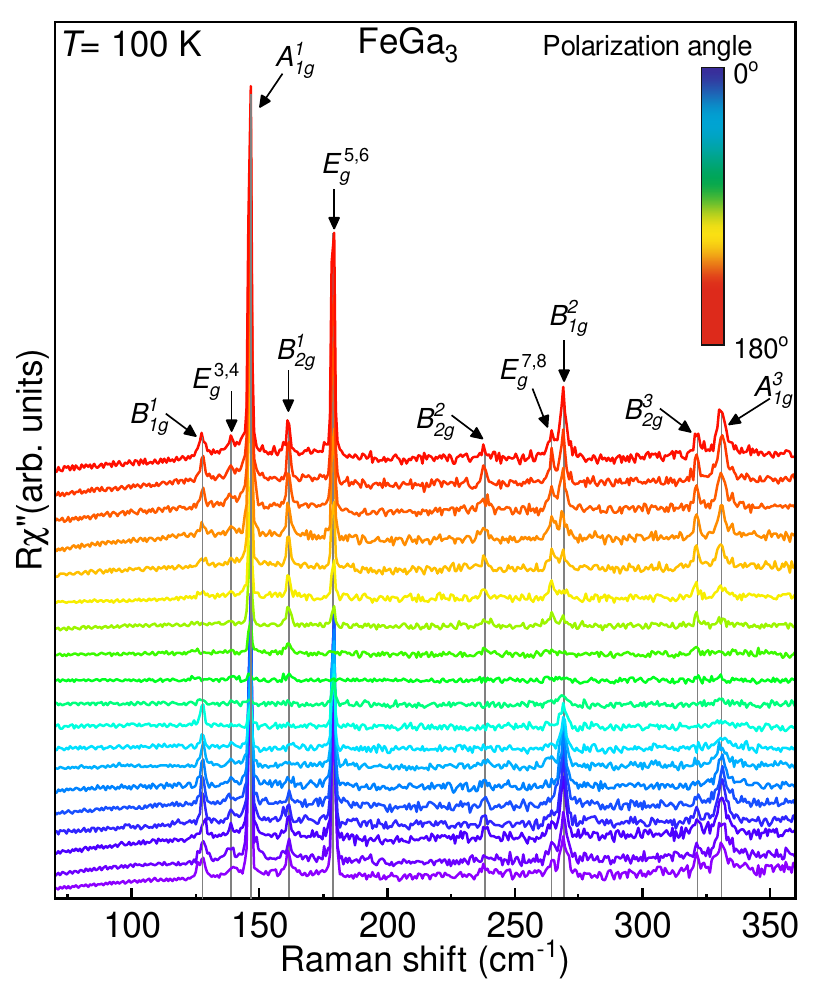}
  \caption{Polarization-dependent Raman spectra of FeGa$_3$ single crystals. Measurements were performed with measuring step of 10$^\circ$ at temperature $T$ = 100\,K.}
 \label{Fig:Raman_New}
\end{figure}

Raman spectra of FeGa$_3$ single crystals, measured from the (011)-plane of the sample, at temperature $T=100$~K, for polarization angles within the range of 0$^\circ$ and 180$^\circ$ are presented in Fig.~\ref{Fig:Raman_New}.
We have identified 10 out of 12 Raman active modes. The assignation of modes was done in accordance with DFT calculations and polarization measurements. Peaks that exhibit the same polarization dependence were assigned with the same symmetry. Consequently, peaks at 146.58 and 331.80 \wn were assigned as $A_{1g}$ and peaks at 127.99 and 269.98 \wn were assigned as $B_{1g}$. The phonon lines at 138.96, 179.01 and 264.40 \wn are assigned as $E_{g}$, whereas modes at 161.67, 238.27 and 321.43 \wn correspond to the  $B_{2g}$ symmetry modes. The full list of measured phonon frequencies, along with their calculated values, is shown in Table~\ref{table2}.

There is a very good agreement between experimental and calculated phonon frequencies, with a discrepancy of less than 8\%. A close match in experimental and theoretical results is not surprising knowing that the investigated material is semiconducting, with moderate electronic correlations. All of the observed phonon lines are sharp, with the full-width at half maximum (FWHM) $\sim$ 2 cm$^{-1}$,
and weakly temperature dependent. This indicates a good crystallinity of the sample and absence of a metal-insulator transition or magnetic ordering.

\begin{table}[t!]
\caption{Experimental Raman frequencies measured at 100\,K and the corresponding values calculated within DFPT.
\vspace*{0.02cm}
}
\label{table2}
\centering

\begin{tabular}{l l l } 
\hline \hline 
Irred.~rep. & Exp. ($\mathrm{cm}^{-1}$) & Calc. ($\mathrm{cm}^{-1}$) \\ [0.5ex] 
\hline 



$E^{1,2}_g$ &  - & 86.81 \\ 
$B^{1}_{1g}$ &  127.99 & 125.52 \\ 
$E^{3,4}_g$ &  138.96 & 139.06 \\ 
$A^{1}_{1g}$ &  146.58 & 145.89 \\ 
$B^{1}_{2g}$ & 161.67 & 161.51 \\ 
$E^{5,6}_g$ &  179.01 & 165.09 \\ 
$A^{2}_{1g}$ &  - & 180.12 \\ 
$B^{2}_{2g}$ &  238.27 & 239.62 \\ 
$E^{7,8}_g$ &  264.40 & 258.94 \\ 
$B^{2}_{1g}$ &  269.28 & 262.96 \\ 
$B^{3}_{2g}$ &   321.43& 318.53 \\ 
$A^{3}_{1g}$ &  331.80 & 322.41 \\ [1ex] 
\hline
\end{tabular}

\end{table}

\subsection{M\"ossbauer spectra}

\begin{figure}[t]
 \centering
  \includegraphics[width=\columnwidth]{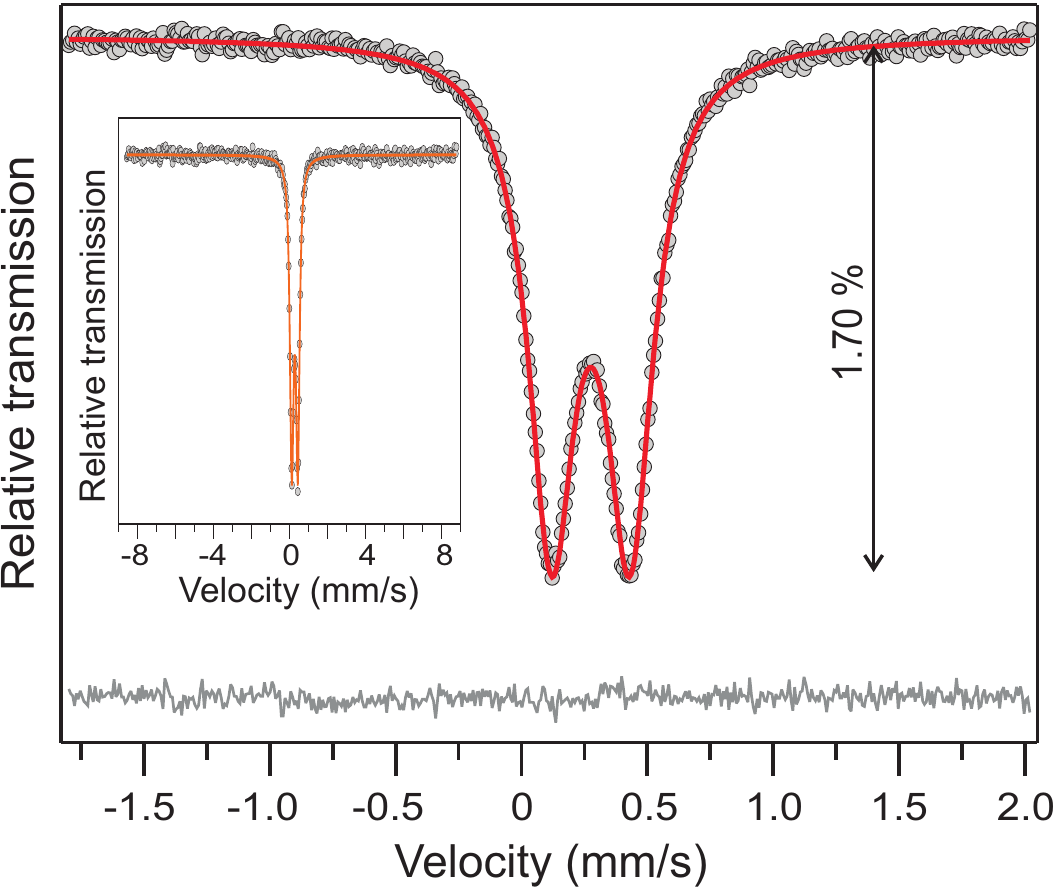}
  \caption{Room temperature $^{57}$Fe-M\"ossbauer spectra of the FeGa$_3$ sample recorded in the low-velocity range. Experimental data are presented by the solid circles and the fit is given by the red solid line. The difference ($\mathrm{calc}-\mathrm{exp}$) is shown by the dark gray line at the bottom of the figure. The vertical arrow denotes the relative position of the lowest experimental point with respect to the background (relative absorption of 1.70\%). The absolute difference is less than 0.05\%. The inset shows the room temperature spectrum of the FeGa$_3$ sample recorded within the high-velocity range. The orange line is just a guide for the eye. 
}
 \label{Fig:Moss}
\end{figure}

The $^{57}$Fe-M\"ossbauer spectroscopy was used to investigate quality and ordering of the prepared sample and to check for the presence of Fe-based impurity phases. The $^{57}$Fe-M\"ossbauer spectra of the FeGa$_3$ are presented in Fig.~\ref{Fig:Moss}. The spectrum recorded in the low-velocity range showed two absorption lines (doublet). In the spectrum recorded in the high-velocity range, beside the observed doublet, there is no indication of the magnetic hyperfine splitting.
The thickness corrected FeGa$_3$ spectrum recorded in the low-velocity range was fitted with one Lorentzian-shaped doublet using the Recoil program \cite{Lagarec_1998}. The obtained M\"ossbauer parameters for the measured doublet are: center shift $CS = 0.28$ mm/s, quadrupole splitting $\Delta = 0.31$ mm/s, and FWHM of the Lorentzian lines is 0.22 mm/s. The obtained results closely match the hyperfine parameters for FeGa$_3$ from the literature \cite{Whittle_1980,Dezsi_1998,Tsujii_2008, Hearne_2018, Mondal_2018}.
A very small broadening of the resonance lines observed in the experiment is a strong indication that the sample is very well ordered and of high crystallinity. 

\section{Conclusions}

In summary, we have performed optical, Raman, and M\"ossbauer spectroscopy measurements of a narrow-gap semiconductor FeGa$_3$, along with DFT band structure and vibrational frequencies calculation. We find that the optical conductivity decreases for frequencies below $\hbar \omega \sim 0.9$~eV consistent with the direct band gap observed in DFT calculations. Our most important finding is the appearance of the optical spectral weight below the charge (indirect) gap $E_g \approx 0.4$~eV. At room temperature the spectral weight below $E_g$ diminishes as compared to the one at $T=25$~K. Therefore, we conclude that this spectral weight originates from the impurities and not from the correlation effects.
Interestingly, we do not find signatures of the impurities in the vibrational spectra. Both the infrared and Raman lines are very narrow, as well as the M\"ossbauer spectral lines, which implies a good crystallinity of the sample. The calculated vibrational frequencies are in a very good agreement with the measurements, which indicates that the electronic correlations in FeGa$_3$ are not strong, in line with previous studies.

\acknowledgments

A.\,U. acknowledges fruitful discussions with V.\,Iva\-novski.
C.\,M. acknowledges funding from the Research Honors Program at Ramapo College of New Jersey.
M.\,O., S.\,Dj.-M., P.\,M., N.\,L.~and D.\,T.~acknowledge funding provided by the Institute of Physics Belgrade, through a grant by the Ministry of Science, Technological Development and Innovation of the Republic of Serbia. A.\,U.~acknowledges support provided by the Vin\v ca Institute of Nuclear Sciences, through an agreement with the Ministry of Science, Technological Development and Innovation of the Republic of Serbia.
C.\,P.~acknowledges support by the U.S.~Department
of Energy, Basic Energy Sciences, Division
of Materials Science and Engineering, under Contract No.~DE-SC0012704 (BNL), and W.\,R.~was supported by the National Natural Science
Foundation of China under Grants No. 51671192 and No.
51531008 (Shenyang).
Numerical simulations were performed on the PARADOX supercomputing facility at the Scientific Computing Laboratory, National Center of Excellence for the Study of Complex Systems, Institute of Physics Belgrade.

%

\end{document}